\documentclass[12pt]{article}
\usepackage{a4}
\usepackage{bm}
\usepackage[dvips]{graphicx}

\def\pz{\phantom{0}}
\def\ds{{\rm D}^-_{\rm s}}
\def\dsf{{\rm D}_{\rm s}}
\def\dss{{\rm D}^{\star -}_{\rm s}}
\def\dssf{{\rm D}^{\star}_{\rm s}}
\def\Z{{\rm Z}}
\def\fds{{\rm f}_{\rm D_{\rm s}}}
\def\mds{m_{\rm D_{\rm s}}}
\def\tds{\tau_{\rm D_{\rm s}}}
\def\dst{\ds\to\tau^-\bar{\nu}_\tau}
\def\dstg{\gamma\ds\to\gamma\tau^-\bar{\nu}_\tau}
\def\dsm{\ds\to\mu^-\bar{\nu}_\mu}
\def\dsmg{\gamma\ds\to\gamma\mu^-\bar{\nu}_\mu}
\def\ccbar{\mbox{c}\overline{\mbox{c}}}
\def\csbar{\mbox{c}\overline{\mbox{s}}}
\def\bbbar{\mbox{b}\overline{\mbox{b}}}
\def\ee{\mbox{e}^+\mbox{e}^-}

\def\tmu{\tau^-\to\mu^-\bar{\nu_\mu}\nu_\tau}
\def\tel{\tau^-\to e^-\bar{\nu_e}\nu_\tau}
\def\alp{\alpha_{\rm lep,\rm D_{\rm s}}}
\def\plep{p_{\rm lep,\rm D_{\rm s}}}

\def\efftmu{0.9\%}
\def\efftel{0.7\%}
\def\effmu{0.6\%}
\def\resultbr{\left(7.0\pm 2.1(\mbox{stat})\pm 2.0(\mbox{syst})\right)\%}
\def\resultf{\left(286\pm 44(\mbox{stat})\pm 41(\mbox{syst})\right)\mbox{MeV}}
\def\numcan{22.5 \pm 6.9}

\newcommand {\downto}
        {\mbox{ \begin{picture}(14,10)
                   \put(0,10){\line(0,-1){5.0}}
                   \put(2,5){\oval(4,4)[bl]}
                   \put(1,0){\makebox(0,0)[bl]{$\rightarrow$}}
                \end{picture} }}

\begin{document}


\begin{titlepage}
\begin{center}{\large   EUROPEAN ORGANIZATION FOR NUCLEAR RESEARCH
}\end{center}\bigskip
\begin{flushright}
       CERN-EP-2001-019   \\ 6 March 2001
\end{flushright}
\bigskip\bigskip\bigskip\bigskip\bigskip
\begin{center}{\huge\bf  Measurement of the Branching Ratio for $\bm{\dst}$ Decays 
}\end{center}\bigskip\bigskip
\begin{center}{\LARGE The OPAL Collaboration
}\end{center}\bigskip\bigskip
\bigskip\begin{center}{\large  Abstract}\end{center}
\noindent
Using about 3.9 million hadronic Z decays 
from ${\rm e}^+{\rm e}^-$ collisions recorded by
the OPAL detector at LEP 
at centre-of-mass energies $\sqrt{s}\approx M_{\rm Z}$,
the branching ratio for the 
decay $\dst$ has been measured to be
$$\mbox{BR}(\dst)=\resultbr.$$
This result can be used to derive the
decay constant of the $\ds$ meson:
$$\fds = \resultf.$$
\bigskip\bigskip\bigskip\bigskip
\bigskip\bigskip
\begin{center}{\large
(submitted to Physics Letters B)
}\end{center}
\end{titlepage}
\begin{center}{\Large        The OPAL Collaboration
}\end{center}\bigskip
\begin{center}{
G.\thinspace Abbiendi$^{  2}$,
C.\thinspace Ainsley$^{  5}$,
P.F.\thinspace {\AA}kesson$^{  3}$,
G.\thinspace Alexander$^{ 22}$,
J.\thinspace Allison$^{ 16}$,
G.\thinspace Anagnostou$^{  1}$,
K.J.\thinspace Anderson$^{  9}$,
S.\thinspace Arcelli$^{ 17}$,
S.\thinspace Asai$^{ 23}$,
D.\thinspace Axen$^{ 27}$
% ...
}\end{center}\bigskip
$^{  1}$School of Physics and Astronomy, University of Birmingham,
Birmingham B15 2TT, UK
\newline
$^{  2}$Dipartimento di Fisica dell' Universit\`a di Bologna and INFN,
I-40126 Bologna, Italy
\newline
$^{  3}$Physikalisches Institut, Universit\"at Bonn,
D-53115 Bonn, Germany
\newline
% ...
$^{  5}$Cavendish Laboratory, Cambridge CB3 0HE, UK
\newline
% ...
$^{  9}$Enrico Fermi Institute and Department of Physics,
University of Chicago, Chicago IL 60637, USA
\newline
% ...
$^{ 16}$Department of Physics, Schuster Laboratory, The University,
Manchester M13 9PL, UK
\newline
$^{ 17}$Department of Physics, University of Maryland,
College Park, MD 20742, USA
\newline
% ...
$^{ 22}$Department of Physics and Astronomy, Tel Aviv University,
Tel Aviv 69978, Israel
\newline
$^{ 23}$International Centre for Elementary Particle Physics and
Department of Physics, University of Tokyo, Tokyo 113-0033, and
Kobe University, Kobe 657-8501, Japan
\newline
% ...
$^{ 27}$University of British Columbia, Department of Physics,
Vancouver BC V6T 1Z1, Canada
\newline
% ...
Tel Aviv 69978, Israel.
\section{Introduction}
\label{sec-intro}%
The branching ratio of the purely leptonic $\ds\to 
\ell^-\bar{\nu_\ell}$ decay\footnote{Charge conjugate decays are
implied throughout the paper.}
can be calculated~\cite{bib-pot} using
\begin{equation}
\label{eq-th}
\mbox{BR}(\ds\to \ell^-\bar{\nu}_\ell)=
\frac{G_{\mbox{\scriptsize F}}^2 }{8\pi} \mds 
m_\ell^2\left(1-\frac{m_{\ell}^2}{\mds^2}\right)^2|V_{\rm cs}|^2\tds 
\fds^2,
\label{eq1}
\end{equation}
where $\mds$ is the mass and $\tds$ the lifetime of the $\ds$ meson, 
$\fds$ the $\ds$ decay constant and
$V_{\rm cs}$ the corresponding CKM matrix element. 
$G_{\mbox{\scriptsize F}}$ denotes the Fermi coupling constant and
$m_{\ell}$ the mass of the lepton.

Several models for the calculation of the decay constant $\fds$ 
exist: potential models predict  $\fds$ in the range 
from 129 MeV to 356 MeV~\cite{bib-pot}, QCD sum rule models 
predict $\fds=\left(231\pm 24\right)$~MeV~\cite{bib-sum} 
and lattice QCD calculations predict $\fds = (240^{+30}_{-25})$~MeV~\cite{bib-latqcd}.

The extraction of CKM matrix elements from 
$\mbox{B}^0-\overline{\mbox{B}}^0$ oscillation measurements
relies on these theoretical models for calculation of the decay constant for 
B mesons, $f_{{\rm B}}$, since a measurement of $f_{{\rm B}}$ from
B$^-\to\ell^-\bar{\nu}_{\ell}$ decays is currently not feasible.
It is therefore important to measure $\fds$ to
test the theoretical models used in the $f_{{\rm B}}$ calculation.
Measurements of $\fds$ in leptonic $\ds$ decays have been
published by WA75~\cite{bib-wa75}, BES~\cite{bib-bes}, 
E653~\cite{bib-e653}, L3~\cite{bib-l3ds}, 
CLEO~\cite{bib-cleo}, and BEATRICE~\cite{bib-beatrice}. 
The measured values lie between 190~MeV and 430~MeV. The current
world average is $280\pm48$~MeV~\cite{bib-pdg}.

In this paper, we present a measurement of BR($\dst$) and 
$\fds$ based on reconstruction of the decay sequence
\begin{tabbing}
\hspace{10mm} \= \hspace{33mm} \= \hspace{7mm} \= \hspace{3mm} \=
\hspace{7mm} \= \hspace{40mm}  \=  \kill 
 \>$\ee\to\Z\to\ccbar\rightarrow$\>$\dss X$ \>                  \>           \>\> 
 \\
 \>                        \>$\downto$\> $\gamma$\> $\ds$\>\> 
 \\
 \>                        \>         \>                  \>$\downto$  \> 
$\tau\bar{\nu}_{\tau}$ \>  \\
 \>                        \>         \>                  \>\> $\downto 
{\ell}^-\bar{\nu}_{\ell}\nu_\tau\;\; (\ell=e,\mu). $\\
\end{tabbing}
\begin{flushright} \vspace{-1.3cm} (2) \end{flushright}
\setcounter{equation}{2}

Only $\dst$ events from $\Z\to {\rm c}\bar{\rm c}$ decays are considered,
since a measurement of BR($\dst$) in $\Z\to {\rm b}\bar{\rm b}$ events is 
systematically limited by the large uncertainty 
on the production rate of $\ds$ mesons in $\Z\to {\rm b}\bar{\rm b}$ events.

Hadronic $\tau$ decays are difficult to distinguish from 
background and therefore only $\tau$ decays into electrons or muons are 
used. 
Since the $\ds$ mass cannot be reconstructed from a single
particle in the final state, a neural network is trained 
on a preselected sample of hadronic Z events with one identified
electron or muon, requiring the kinematics to be consistent 
with $\dst\to\ell^-\bar{\nu}_{\ell}\nu_\tau\bar{\nu}_\tau$ decays.

In the last step of the analysis $\dss\to\gamma\ds$ decays are 
reconstructed in this $\dst$ enhanced sample by forming
the invariant mass of the photon and the $\ds$ candidate. 
This reduces the 
dependence on the Monte Carlo simulation of the background and
increases the purity of the $\ds$ sample.

Since the $\ds\to\ell^-\bar{\nu_\ell}$ decay is helicity suppressed, 
the $\tau$ channel 
has the largest branching ratio of all leptonic channels.
Eq.~\ref{eq1} predicts the branching ratio into electrons to be negligible,
${\rm BR}(\ds\to{\rm e^-}\bar{\nu_{\rm e}})/{\rm BR}(\dst)<10^{-5}$, due
to the factor $m_{\ell}^2$,
whereas the branching ratio into muons, BR($\dsm$), is expected to be 
sizeable, ${\rm BR}(\dsm)/{\rm BR}(\dst)=0.103$. 
Therefore the decay $\dsm$ is included in the
signal definition and the final result is corrected for this contribution.

\section{Detector, data sample and event preselection}
The OPAL detector is described in detail elsewhere~\cite{bib-opal}. 
Tracking of charged particles is performed by a central detector,
consisting of a silicon microvertex detector, a vertex chamber, a jet chamber
and $z$-chambers\,\footnote{A right handed coordinate system is used, with
positive $z$ along the $\mathrm{e}^-$ beam direction and $x$ pointing
towards the centre of the LEP ring. The polar and azimuthal angles are
denoted by $\theta$ and $\phi$, and 
the origin is taken to be the centre of the detector.}.
The central detector is inside a
solenoid, which provides a uniform axial magnetic field of 0.435\,T.
The silicon microvertex detector consists of two layers of
silicon strip detectors; for most of the data used in this paper,
the inner layer covered a polar angle range of $|\cos\theta |<0.83$ and
the outer layer covers $| \cos \theta |< 0.77$, with
an extended coverage for the data taken after the year 1996.
The vertex chamber is a precision drift chamber
which covers the range $|\cos \theta | < 0.95$.
The jet chamber is
a large-volume drift chamber, 4.0~m long and 3.7~m in diameter,
providing both tracking and ionization energy loss (d$E$/d$x$) information.
The $z$-chambers provide a precise measurement of the $z$-coordinate
of tracks as they leave the jet chamber in the range
$|\cos \theta | < 0.72$.

The coil is surrounded by a time-of-flight counter array and
a barrel lead-glass electromagnetic calorimeter with a presampler.
Including the endcap electromagnetic calorimeters,
the lead-glass blocks cover the range $| \cos \theta | < 0.98$.
The magnet return yoke is instrumented with streamer tubes
and serves as a hadron calorimeter.
Outside the hadron calorimeter are muon chambers, which
cover 93\% of the full solid angle.

For Monte Carlo studies, event samples have been generated
using JETSET 7.4~\cite{bib-jt74} for multihadronic $\Z$~events 
and KORALZ 4.0 \cite{bib-koralz} for $\tau$ pair events. 
Special signal samples have also been generated using JETSET.
These consist of 
$\Z\to\ccbar\to\ds X$ events with the decay sequence given in Eq.~2. 
The $\dst$ signal is normalised using 
$f(c\to \ds)=0.121$~\cite{bib-frag} and $\mbox{BR}(\dst)=7.0\%$.
Tau polarisation effects are handled by the $\tau$ decay library 
TAUOLA 2.4~\cite{bib-tauola}.

The data sample
used in this analysis consists of about 3.5 million $\Z$ decays recorded
during the period 1991-1995 and an additional 0.4 million $\Z$ events
recorded for detector calibration purposes in 1996-2000. 
Events are only used if the silicon microvertex detector and the
other main detector components relevant
for the analysis were fully operational. 
Hadronic $\Z$ decays are selected based on the number of reconstructed
tracks and the energy deposited in the calorimeter~\cite{bib-mhsel}.
To ensure that the event is well contained within the acceptance of the 
central detector, the polar angle of the thrust axis 
is required to satisfy $|\cos \theta_{\rm T} | <0.8$.

Signal events are characterised by the presence of an electron or a muon from 
$\tau$ decays and large missing energy. 
Electrons and muons are identified using neural 
networks~\cite{bib-elid,bib-muid} which are trained to 
identify leptons with a momentum greater than $2$~GeV.  Only events
with exactly one identified electron or muon are selected. 
Electrons from photon conversions are rejected using a neural network
conversion finder~\cite{bib-elconv}. 

Each event is divided into two hemispheres by the plane perpendicular 
to the thrust axis.
The hemisphere with less visible energy is required to contain the lepton.
This hemisphere is selected to search for $\dst$ decays.
To further enrich the sample in ${\rm c}\bar{\rm c}$ events, a 
loose anti-b tag~\cite{bib-btag} is applied in the hemisphere opposite to
the $\ds$.
Finally, at least 9 tracks are required in an event
to reduce the background from $\Z\to\tau^+\tau^-$ events while
keeping more than $97\%$ of the signal events at this stage of 
the selection.

\section{Reconstruction technique}
\label{sec-dsrec}%
A matching algorithm \cite{bib-mt} is used to avoid double-counting of 
particle momenta in the calorimeters and in the tracking detectors.
The output of the matching algorithm -- referred to as particles --
are tracks and calorimeter clusters.

If the missing energy in the event is only due to the
neutrinos produced in the $\ds$ decay, the energy and momentum of the 
$\ds$ are exactly given by 
\begin{eqnarray}
\vec{P}_{\dsf} &=& -\sum_{i\ne \mbox{\scriptsize lepton}} \vec{p}_{i}\\
E_{\dsf} &=&\sqrt{s}-\sum_{i\ne \mbox{\scriptsize lepton}} E_{i}\; ,
\end{eqnarray}
where $\sqrt{s}$ is the ${\rm e}^+{\rm e}^-$ centre-of-mass energy.
The summation is performed over all particles in the event except the
lepton. 
The resulting mean reconstructed energy of the $\ds$ is 27 GeV which is
slightly larger than the true mean energy of 26 GeV.

Due to detector acceptance and resolution effects this method yields
an energy resolution of $6.5$~GeV
and an angular
resolution of $52$~mrad where the resolution is defined
as the sigma of a single Gaussian fitted to the distribution.  
To further improve the energy resolution, a kinematic
fit is applied in which the energy and the absolute momentum of all particles
(except the lepton) are varied independently from each other (i.e. varying their mass) using the constraint

\begin{equation}
\sqrt{E_{\dsf}^2-P^2_{\dsf}}=M_{\dsf}.
\end{equation}

The $\chi^2$ values calculated from the deviations 
from the experimentally measured values
\begin{equation}
\chi^2=\sum_{i\ne {\rm lepton}}\frac{(E^{\rm fit}_i-E^{\rm meas}_i)^2}
{\sigma^2_{E^{\rm meas}_i}}
\end{equation}
are minimized.
This procedure yields an energy resolution of about $3.0$~GeV.
About $2\%$ of the events are rejected because the kinematic fit does 
not converge.
The efficiency $\epsilon(\dst)$ 
to reconstruct the $\Z\to\ccbar\to\dss X,\; \dss\to\dstg$ 
signal in the $\tmu$ channel or in the $\tel$ channel 
at this stage of the analysis is about $30\%$.

\section{Selection of $\bm{\dst}$ candidates}
In the next part of the analysis a $\dst$ enriched sample is selected
using neural networks. About $52\%$ of the selected
events used as input to the neural networks are expected
to be $\Z\to\bbbar$ events, about $36\%$ $\Z\to\ccbar$ events and
the remaining $12\%$ Z boson decays into light quarks (uds).
The signal contribution is expected to be of the order $1\%$.
For each channel (electron or muon) two neural networks are trained: 
one to separate signal from ${\rm Z}\to{\rm c}\bar{\rm c}$ background events 
and one to distinguish between signal and ${\rm Z}\to{\rm b}\bar{\rm b}$ 
background.

The light-quark background is not used in the training.
Since $\ds\to\tau\nu_{\tau}$ decays from b decays
are not considered signal, they are included in the
${\rm b}\bar{\rm b}$ background.
They constitute about $0.8\%$ of the ${\rm b}\bar{\rm b}$ 
background events used as input to the neural network.

\begin{figure}
\begin{center}
\includegraphics[scale=0.82]{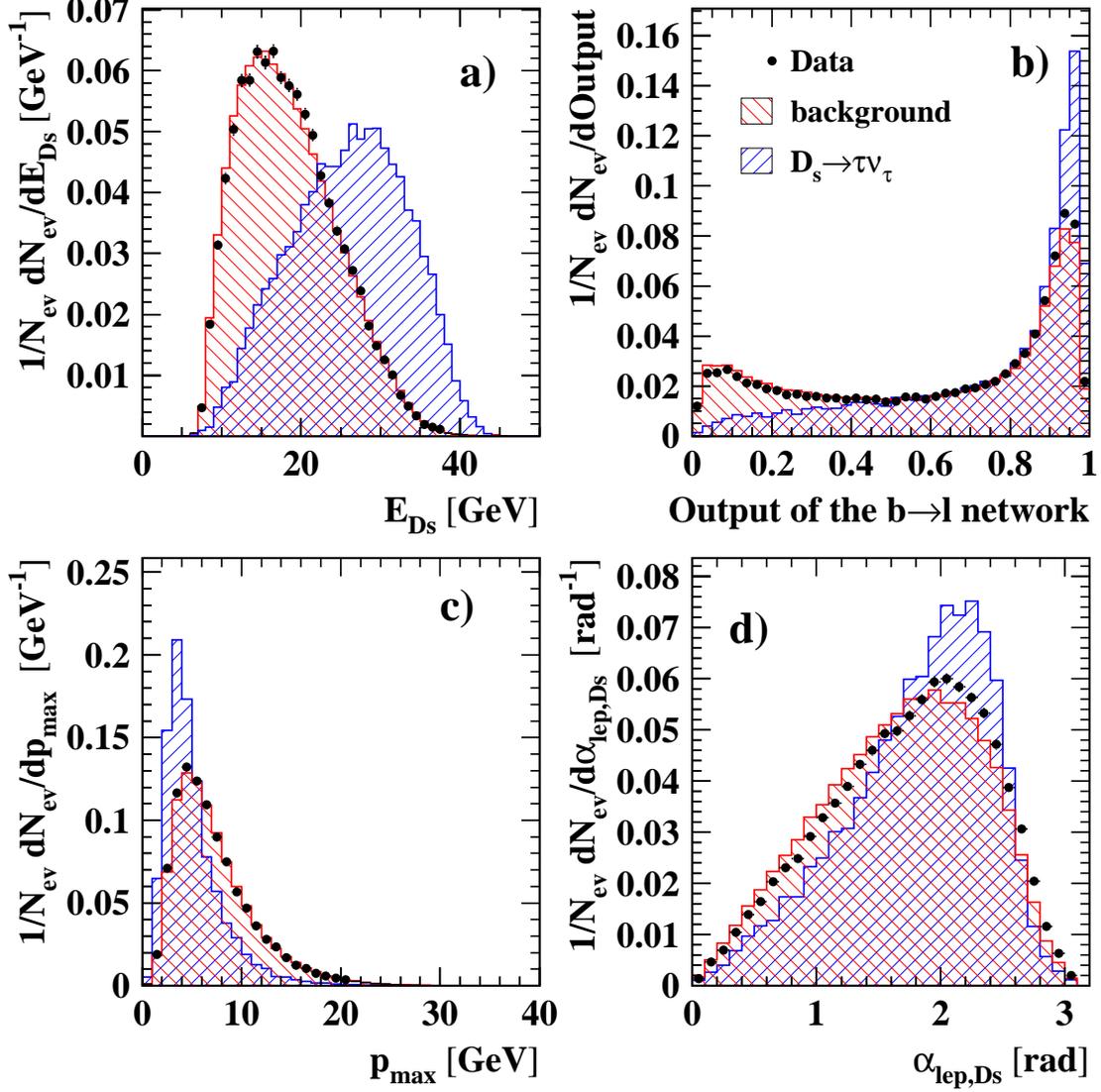}
\end{center}
\caption{\label{input-vars}
Four selected input variables of the neural networks used in the electron 
channel: 
a) reconstructed energy $E_{\dsf}$ of the $\ds$;
b) output of the neural networks 
trained to find ${\rm b}\to \ell$ decays;
c) highest momentum $p_{\rm max}$ of any particle 
with a charge opposite to that of the lepton in the $\ds$ hemisphere.
d) angle $\alp$ between the lepton in the $\ds$ rest 
frame and the $\ds$.
All distributions are normalized to the number of events, $N_{\rm ev}$.
The signal contribution in the data is about $1\%$ at this
stage of the selection.
}
\end{figure}

The following variables are used in all four networks:

\begin{itemize}
\item The reconstructed energy $E_{\dsf}$ of the $\ds$ obtained from the   
kinematic fit (Fig.~\ref{input-vars}a); 
the reconstruction method used for the energy and the momentum of the 
$\ds$ is only valid for purely leptonic decays. 
Semileptonic background decays are expected to have a lower reconstructed 
$\ds$ energy $E_{\dsf}$. 

\item The lepton energy $E_{\rm lep}$;
leptons in light-quark background events
have on average lower energy than in signal events whereas 
leptons in background events from 
${\rm b}\to {\ell}$ 
decays have on average higher energy than in signal events due to the hard 
fragmentation of the b hadron. 

\item The output of two additional neural networks 
trained to find ${\rm b}\to \ell$ 
(Fig.~\ref{input-vars}b) and 
${\rm b}\to {\rm c}\to \ell$ decays~\cite{bib-bt};
leptons originating from signal events have properties more similar 
to leptons from ${\rm b}\to \ell$ decays than from 
${\rm b}\to {\rm c}\to \ell$ decays.

\item The visible invariant mass determined from the tracks and clusters 
and the energy sum in the electromagnetic calorimeter (ECAL), 
both calculated in the $\ds$ hemisphere; 
on average they are lower for signal events due to the energy 
carried by the neutrinos.

\end{itemize}

The choice of input variables is optimized separately for each net. 
The following two variables are only used in the two
neural networks rejecting $\bbbar$ background and 
in the neural network separating the muon channel from
the $\ccbar$ background:

\begin{itemize}

\item The momentum $\plep$ of the lepton in the $\ds$ rest frame
for signal events; it is limited by the mass of the $\ds$ meson
to be $\plep<\mds/2$.
The $\plep$ distribution is smeared by the experimental resolution.
For leptons not originating from $\ds$ decays this restriction does not exist, 
leading to a tail at higher $\plep$.

\item The angle between the direction of the reconstructed $\ds$ and 
the direction of the jet containing the $\ds$; 
in signal events this angle is on average larger than in 
$\Z\to {\rm q}\bar{\rm q}$ background events.
The jets are reconstructed by combining all particles - including the lepton -
using a cone algorithm~\cite{bib-cone}.
The jet direction is then calculated excluding the lepton.

\end{itemize}

Variables sensitive to the flavour
of the event are used in the neural networks
separating signal from $\ccbar$ background:

\begin{itemize}

\item The highest momentum $p_{\rm max}$ of any particle 
with a charge opposite to that of the lepton in the $\ds$ hemisphere
(Fig.~\ref{input-vars}c); 
in signal events this particle should originate from the fragmentation of 
the c quark which produced the $\ds$ meson.
On average, it is therefore expected to 
have less momentum than the highest momentum charged particle in 
$\ccbar$ and light-quark background events.

\item 
The angle $\alp$ between the direction of
the lepton in the $\ds$ rest frame and the direction of the reconstructed 
$\ds$ in the lab frame (Fig.~\ref{input-vars}d);  
for $\ccbar$ background events the $\alp$
distribution is broad and it is peaked around $\pi/2$ while in signal 
events $\alp$ is closer to $\pi$. 
\end{itemize}

In the neural networks which reject
$\bbbar$ background
the following variables are used:

\begin{itemize}
\item The number of tracks and the number of 
clusters in the $\ds$ hemisphere; the number of clusters
and the b likelihood in the opposite hemisphere.
\end{itemize}

Four selected variables which display good signal versus
background separation are shown in Fig.~\ref{input-vars}.
Data and background Monte Carlo distributions are in good
agreement.

\begin{figure}
\begin{center}
\includegraphics[scale=0.82]{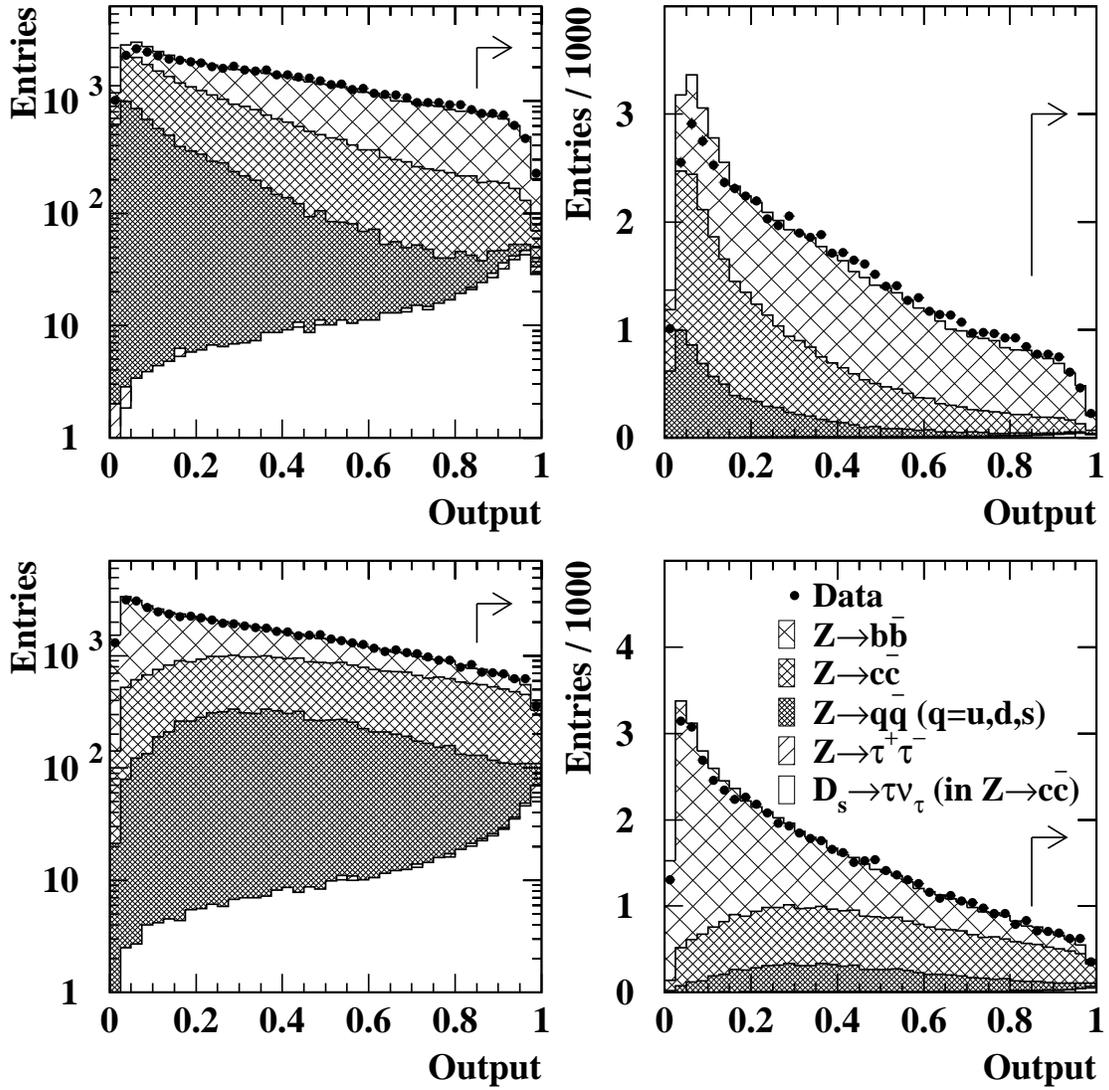}
\end{center}
\caption{\label{muon-nets}%
Muon channel:
a) output of the neural network against 
${\rm Z}\to {\rm c}\bar{\rm c}$ background using a logarithmic scale;
b) a linear scale.  
c) output of the neural network against 
${\rm Z}\to {\rm b}\bar{\rm b}$ background using a logarithmic scale;
d) a linear scale.
The positions of the 
cuts are indicated by arrows.}
\end{figure}

\begin{figure}
\begin{center}
\includegraphics[scale=0.82]{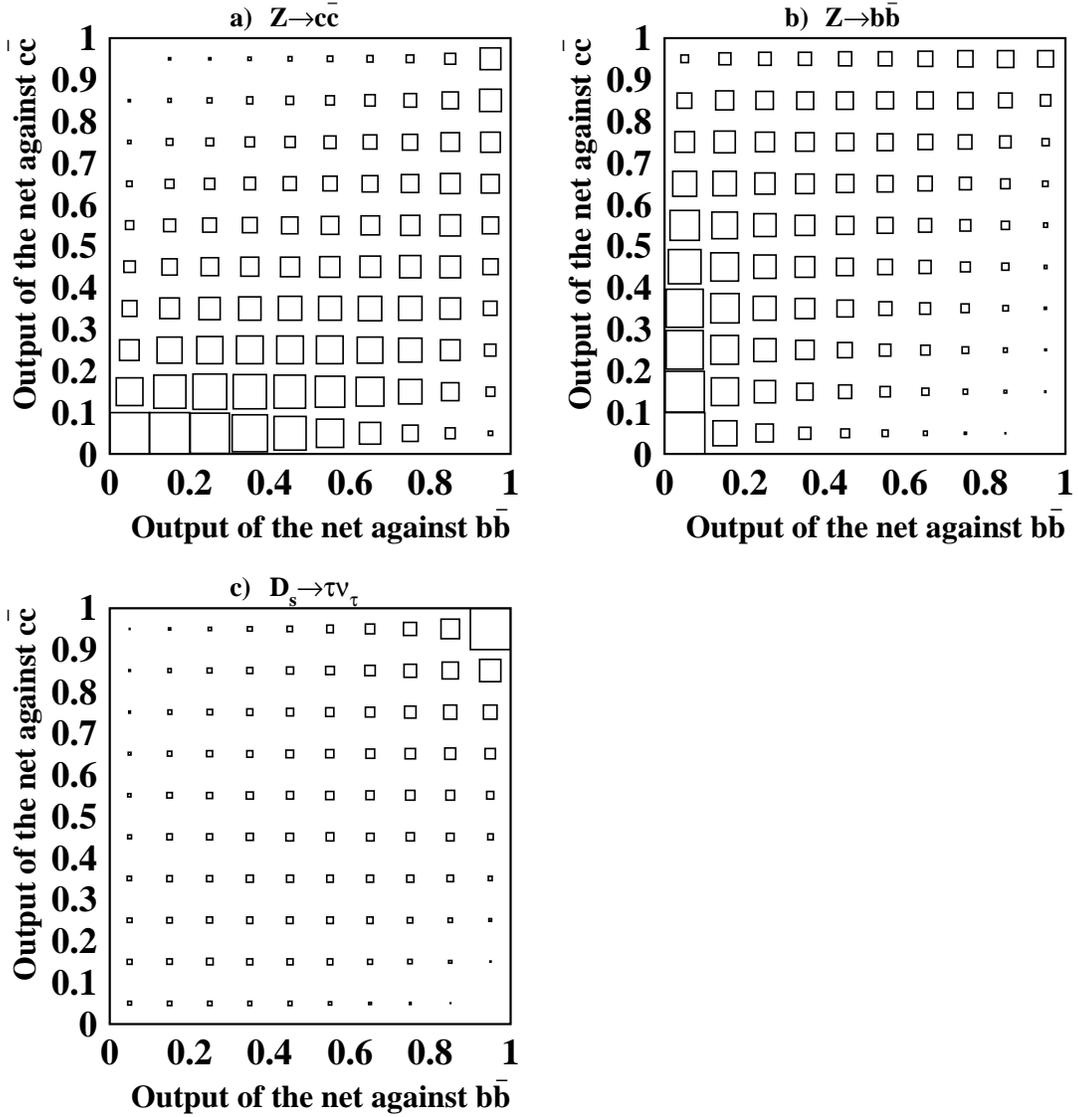}
\end{center}
\caption{\label{fig-twod}
Output of the neural network against 
${\rm Z}\to {\rm c}\bar{\rm c}$ background events versus
the output of the neural network against
${\rm Z}\to {\rm b}\bar{\rm b}$ background events. 
The distributions for the muon and the electron channel have been added
for 
a) ${\rm Z}\to {\rm c}\bar{\rm c}$ background
b) ${\rm Z}\to {\rm b}\bar{\rm b}$ background
and c) signal events in the Monte Carlo.
}
\end{figure}

The output distributions of the neural networks for the muon channel
are shown in  Fig.~\ref{muon-nets}. The output distributions
for the electron channel are similar. 
In Fig.~\ref{fig-twod}, two-dimensional distributions of
the outputs of the neural networks are shown.
A cut at 0.85 on all outputs is applied to 
select a $\dst$ enriched sample with a signal efficiency of
about $\epsilon(\dst)=9\%$.

The discrepancies between data and Monte Carlo are largest
for the neural networks used to reject $Z\to\ccbar$ background
below 0.2, far away from the cut.  
Discrepancies between the 
neural network input distributions of the data and
the Monte Carlo simulation are treated as systematic uncertainties,
as discussed in Section~\ref{sec-sys}.

\section{$\bm{\dss}$ reconstruction}
In the $\dst$ enriched sample, events with photon candidates
in the $\ds$ hemisphere
are used to reconstruct the decay $\dss\to\gamma\ds$.
The $\ds$ signal can then be observed
as a peak in the $\dss$ mass region
of the $\gamma\ds$ invariant mass distribution.

Since the neural networks have been trained to find 
$\dst$ events and not specifically $\dss\to\gamma\ds$ events,
additional cuts on the $\dst$ enriched sample are
required to reduce the background in 
the $\gamma\ds$ invariant mass distribution:
\begin{itemize} 
\item 
The b likelihood as given by the b tagging algorithm~\cite{bib-btag} 
in the $\ds$ hemisphere has to be less than 0.5 to further suppress  
$\bbbar$ background.
\item 
Using energy and momentum conservation, the missing energy in the hemisphere 
is reconstructed from the visible energy $E_{\rm vis}^{\rm hemi}$ in the 
hemisphere, the invariant mass of all particles in the hemisphere,
$M_{\rm hemi}$, and in the opposite hemisphere, $M_{\rm opp}$, via
the relation:
\begin{equation}
E_{\rm miss}^{\rm hemi} = 
\frac{\sqrt{s}}{2}+\frac{M_{\rm hemi}^2-M_{\rm opp}^2}{2 \sqrt{s}}-
E_{\rm vis}^{\rm hemi}. 
\label{eq-emiss}
\end{equation} 
The missing energy $E_{\rm miss}^{\rm hemi}$ has to be larger than 15 GeV.
\end{itemize}
These two cuts reduce the efficiency for the signal to $8\%$.

\begin{figure}
\begin{center}
\includegraphics[scale=0.82]{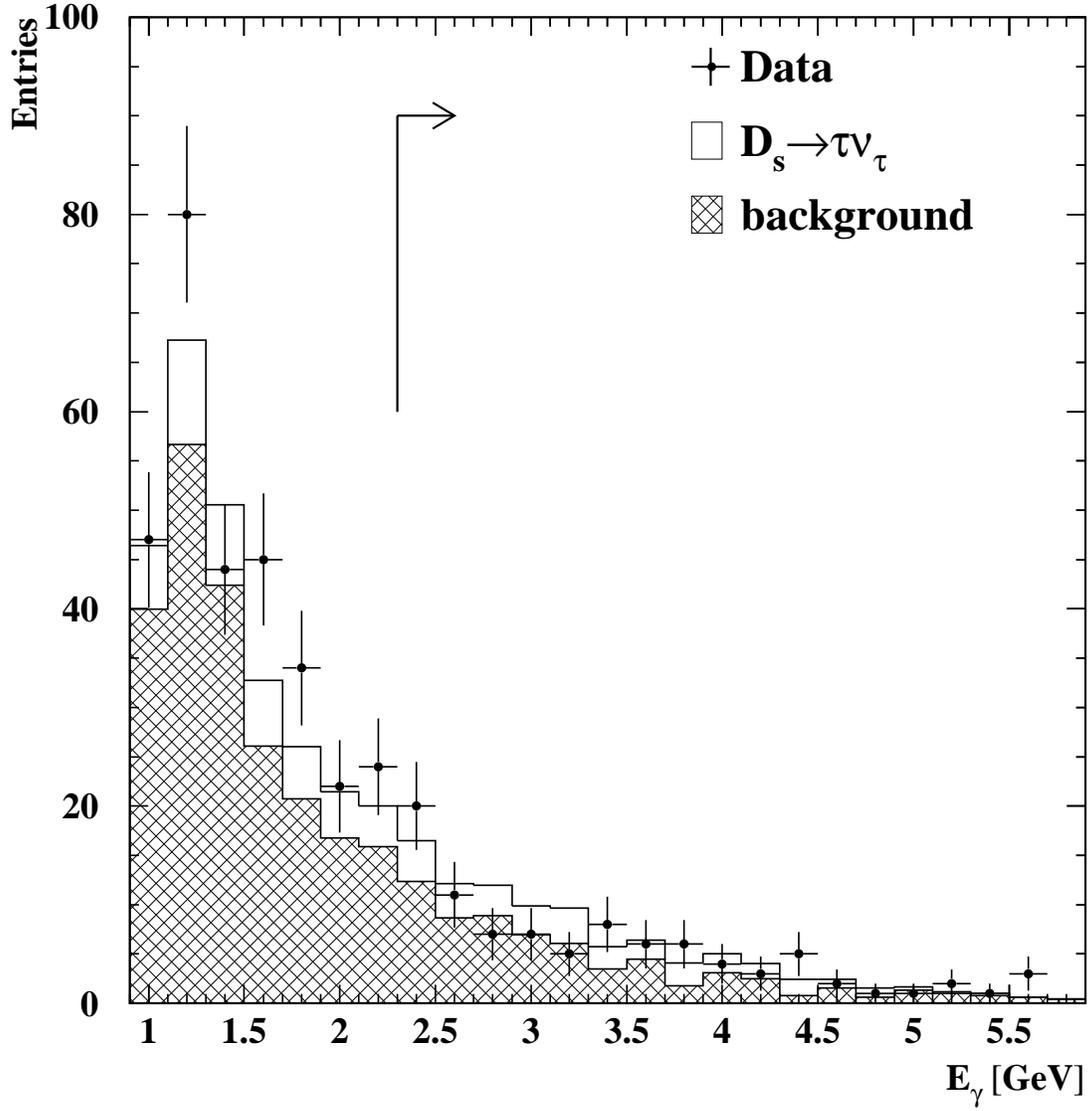}
\end{center}
\caption{\label{fig-egamma}%
Distribution of the photon energy $E_{\gamma}$ for data compared
to the Monte Carlo simulation. 
The position of the 
cut on $E_{\gamma}$ is indicated by an arrow.}
\end{figure}

The photon is found using information from the electromagnetic calorimeter as
described in~\cite{bib-ppefinder}. This method 
assigns a weight to each photon candidate corresponding to the probability 
for it to stem from a real photon.
To accept a photon candidate, this weight has to be larger than $0.6$.
Only events with exactly one such photon candidate in the $\ds$ hemisphere
are accepted. 
The distribution of the photon energy $E_{\gamma}$ after all 
previously defined selection cuts is shown
in Fig.~\ref{fig-egamma}. 
Data and Monte Carlo simulation are in reasonable agreement.
Finally, $E_{\gamma}$ is required to be greater than $2.3$~GeV.
For smaller photon energies the shapes of the 
invariant mass distributions of the photon and the $\ds$ candidate 
become similar for background and signal. 
The energy resolution for photons with $E_{\gamma}>2.3$~GeV as determined by
the Monte Carlo is about $300$~MeV and the angular resolution is about 5~mrad.

\section{Results}
\begin{figure}
\begin{center}
\includegraphics[scale=0.82]{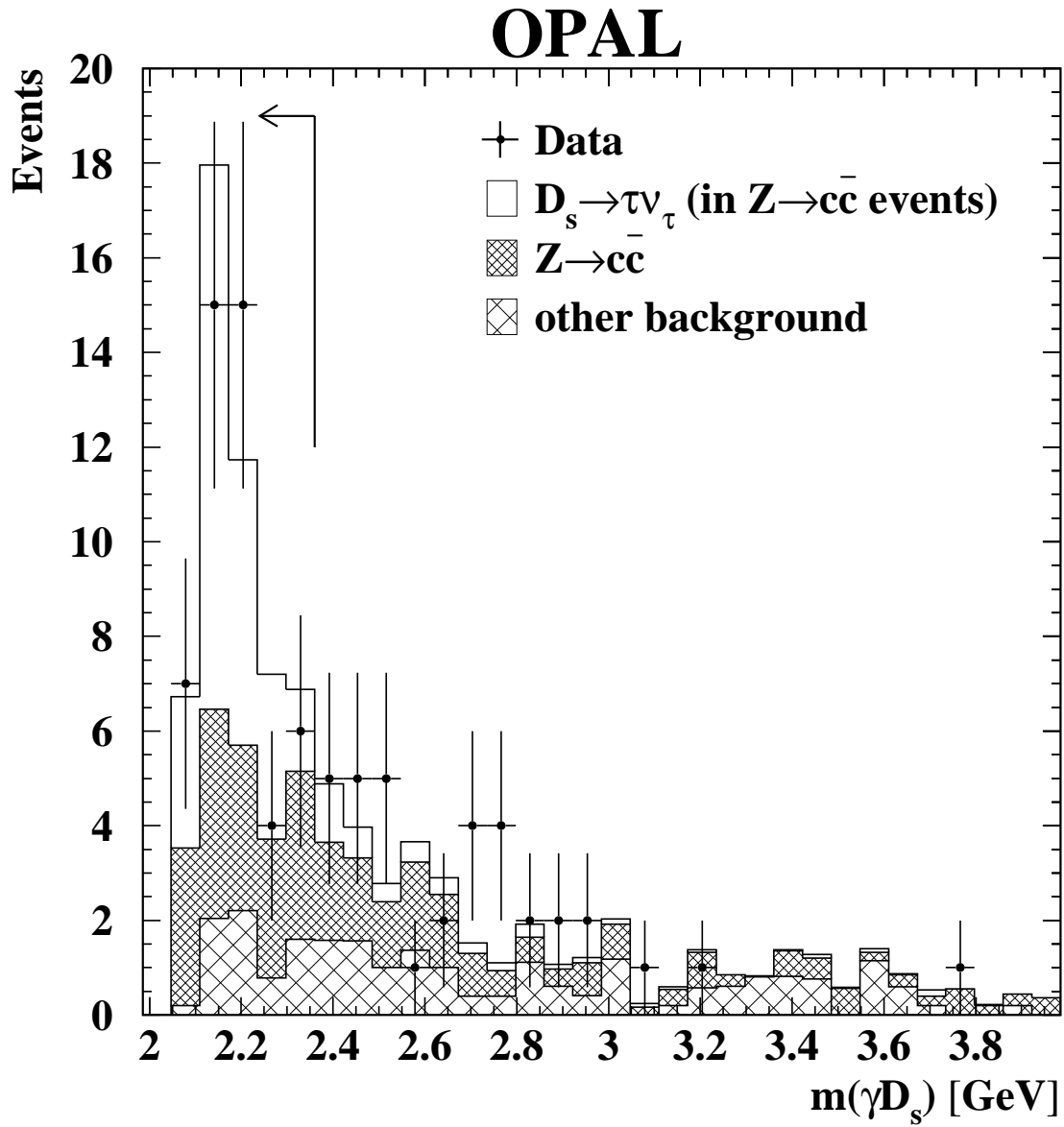}
\end{center}
\caption{\label{fig-res} 
Invariant mass $m(\gamma\ds)$ of the photon and
the $\ds$ candidate for the events satisfying all selection criteria.
The contributions to the Monte Carlo distribution from
the signal and from the different sources of background  
are shown separately. The signal region is indicated by an arrow.}
\end{figure}

The distribution of the invariant mass $m(\gamma\ds)$
of the photon and the $\ds$ candidate for the events
satisfying all selection criteria is shown in Fig.~\ref{fig-res}.
In the signal region, $m(\gamma\ds)<2.36$~GeV, there are
$24.5\pm2.8$ background events predicted by the Monte Carlo.
The number of background events is determined by requiring the
expected number of Monte Carlo events to be identical to
the number of data events after the lepton
identification cuts described in Section~2.

The most important branching ratios have been adjusted in the Monte Carlo 
using the values in~\cite{bib-pdg}.
The uncertainty on the background 
is due to the limited number of Monte Carlo events.  
This number is subtracted from the data which yields
$\numcan$ signal events. The uncertainty is the statistical uncertainty
of the data.

The efficiency to reconstruct $\Z\to\ccbar\to\dss X,\; \dss\to\dstg$ events
in the $\tmu$ channel is $\efftmu$ 
and in the $\tel$ channel $\efftel$. 
If the $\ds$ decays directly into muons via 
$\ds\to\mu^-\bar{\nu}_{\mu}$ the efficiency is $\effmu$.

The branching ratio $\mbox{BR}(\dst)$ is extracted using
\begin{eqnarray}
\mbox{BR}(\dst) = \frac{N_{\mbox{\scriptsize cand}}}{
2 N_{\Z}\cdot R_{\mbox{c}} \cdot f(c\to \ds) \cdot
P_V(\dssf,\dsf)
\cdot\mbox{BR}(\dss\to\gamma\ds)}
 \nonumber \\
\times\frac{1}{\mbox{BR}(\tau\to l\bar{\nu}_l\nu_\tau)\cdot\epsilon
(\dst) + \frac{{\rm BR}(\dsm)}{{\rm BR}(\dst)} 
\cdot\epsilon(\dsm)},\quad
\end{eqnarray}
where
$N_{\rm cand}$ is the number of background-subtracted candidates
in the signal region, 
$N_{\Z}$ the number of $\Z$ decays,
$R_{\rm c}=0.1671\pm0.0048$~\cite{bib-frag} 
the partial width of the $\Z$ decaying into a pair of 
charm quarks, $f(c\to \ds)=0.121\pm0.025$~\cite{bib-frag} the production rate
of $\ds$ mesons in charm jets,
$\epsilon (\dst)$ the efficiency for the signal and $\epsilon (\dsm)$ 
the efficiency for reconstructing $\dss\to\dsmg$ decays.
As discussed in Section~\ref{sec-intro}, we use
${\rm BR}(\dsm)/{\rm BR}(\dst)=0.103$. 

$P_V(\dssf,\dsf)$ is the ratio of $\csbar$ mesons produced in a 
vector state ($\dssf$) with respect to the
sum of the pseudoscalar ($\dsf$) and vector states.
For non-strange D mesons, $P_V({\rm D}^{\star},{\rm D})$ has been measured 
by ALEPH~\cite{bib-apv},
DELPHI~\cite{bib-dpv} and OPAL~\cite{bib-opv}. The averaged
value is $P_V({\rm D}^{\star},{\rm D})=0.61\pm0.03$~\cite{bib-tpv}.
To extrapolate this ratio to $\dsf$ mesons, the effect
of the decays of $L=1$ D$^{\star\star}$ resonances and quark
mass effects need to be taken into account. 
D$^{\star\star}$ resonances contribute only in the case of non-strange mesons.
This effect 
was estimated by OPAL
to be smaller than the experimental uncertainty~\cite{bib-opv}
 and is therefore neglected.
Applying the correction factor for quark mass effects from~\cite{bib-tpv}
yields $P_V(\dssf,\dsf)=0.64\pm0.05$ where the full size of the
correction is included in the uncertainty. This value
is consistent with the ALEPH measurement of 
$P_V(\dssf,\dsf)=0.60\pm0.19$~\cite{bib-apv}.

Using $P_V(\dssf,\dsf)=0.64\pm0.05$ and
the values given in Table~\ref{br-table} for the branching ratios
yields 
\begin{equation}
\mbox{BR}(\dst)=\left(7.0 \pm 2.1 (\mbox{stat})\right)\%.
\end{equation}

\section{Systematic Uncertainties}
\label{sec-sys}
Systematic uncertainties arise from the uncertainties in the branching ratios, 
the Monte Carlo modelling, selection
efficiencies and the detector resolution.
The resulting systematic uncertainties 
are summarised in Table \ref{tab-syst} and
described in more detail below.

\begin{description}
  
\item[External sources:] 
 The external values used in the calculation of
 the branching ratios are each varied within their uncertainties.

\item[Monte Carlo statistics] 
 The uncertainty on the background rate and on the efficiencies 
 $\epsilon (\dst)$ and $\epsilon (\dsm)$ 
 due to the limited number of Monte Carlo events is counted as 
 systematic uncertainty.

\item[Background:]
 To account for uncertainties in the determination of the background rate, 
 the number of background events is also calculated
 using the sideband of the $m(\gamma\ds)$ distribution, 
 defined by $m(\gamma\ds)>2.4$ GeV. The difference between the standard
 analysis and the predicted background rate using the sideband 
 is 0.1 background events in the signal region.
 This difference is taken as systematic uncertainty. 

\item[Background composition:] 
 About $55\%$ of the background is due to combinations
 of $\ds$ candidates with photons which do not originate
 from the same decay.
 The remaining background mainly consists of 
 ${\rm D}^{\star-}\to\gamma {\rm D}^-$ and 
 ${\rm D}^{\star 0}\to\gamma {\rm D}^0$ decays. 
 No $\ds\to\tau\nu_{\tau}$ decays from b decays
 are expected in the final sample.
 The most important branching ratios have been 
 varied within their uncertainties using 
 $\mbox{BR}({\rm D}^{\star -}\to\gamma {\rm D^-})=0.016\pm 0.004$,
 $\mbox{BR}({\rm D}^{\star 0}\to\gamma {\rm D^0})=0.381\pm 0.029$,
 $\mbox{BR}({\rm D}^0\to e^- X^+)=0.0675\pm 0.0029$ and
 $\mbox{BR}({\rm D}^0\to \mu^- X^+)=0.068\pm 0.008$~\cite{bib-pdg}.
 The corresponding change of the result is taken as
 systematic uncertainty. 

\item[Fragmentation:] 
 To determine the effect of uncertainties in the Monte Carlo description 
 of the fragmentation of b and c quarks, the 
 distribution of the scaled hadron energy, 
 $x_{\rm E}=2 E_{\rm h}/\sqrt{s}$,  
 is reweighted within the experimental uncertainties for b quarks,
 $\langle x_{\rm E} \rangle=0.702\pm0.008$,  
 and for c quarks,
 $\langle x_{\rm E} \rangle=0.484\pm 0.008$~\cite{bib-frag}.
 The largest of the variations observed using the fragmentation functions 
 of Peterson et al.,
 Collins and Spiller, and Kartvelishvili et al.~\cite{bib-pet} 
 is taken as systematic uncertainty.

\item[Lepton spectrum:] 
 The exact shape of the lepton momentum spectrum for background events is not
 known. 
 Therefore ${\rm b}\to\ell$, ${\rm b}\to {\rm c}\to\ell$ and 
 ${\rm c}\to\ell$ decays are reweighted to reproduce the lepton momentum 
 spectrum in the rest frame of the b or c hadron as predicted by the
 ACCMM~\cite{bib-ACCMM}, the ISGW~\cite{bib-ISGW} and 
 the ISGW**~\cite{bib-ISGW2} models. 
 The same parameters as in~\cite{bib-frag} are used.
 The largest difference between the results obtained using
 the different models is taken as systematic uncertainty.
  
\item[Tracking resolution:] 
 To take into account uncertainties in the modelling of
 the tracking resolution by the Monte Carlo,
 the reconstructed Monte Carlo track parameters are smeared by 
 $\pm 10\%$~\cite{bib-rb}
 and the analysis is repeated. The largest difference between the results
 is taken as an 
 estimate for this source of systematic uncertainty.
  
\item[Photon energy:] 
 The analysis is redone varying the photon energy scale in the Monte Carlo 
 simulation by $\pm 2\%$~\cite{bib-ppefinder} and the difference
 between the results is taken as systematic uncertainty.
 Furthermore it was checked that the result obtained with the low
 purity sample in the range $1\mbox{~GeV}<E_{\gamma}<2.3\mbox{~GeV}$
 is statistically consistent with the result obtained for 
 $E_{\gamma}>2.3\mbox{~GeV}$.

\item[Lepton identification efficiency:] 
 The electron identification efficiency has been studied in~\cite{bib-effel} 
 and has been found to be modelled correctly within $4\%$.  
 The muon identification efficiency has been studied in~\cite{bib-muid},
 giving an uncertainty of $5\%$.
  
\item[Neural networks:] 
 Each of the input distributions has been compared between data and Monte 
 Carlo simulation. The simulated distributions are reweighted for each input 
 variable in turn to agree with the corresponding data distributions, and the 
 analysis is repeated with the weighted events. The resulting differences in 
 the measured branching ratio are added in quadrature to 
 obtain an estimate of the systematic uncertainty
 due to modelling of the input variables.
 This includes the uncertainty due to the Monte Carlo modelling
 of the missing energy $E_{\rm miss}^{\rm hemi}$ 
 since some of the neural network variables
 are strongly correlated to $E_{\rm miss}^{\rm hemi}$.

\item[b tagging:] 
 The cut on the output of the b-tag in the $\ds$ hemisphere is 
 varied in the Monte Carlo between 0.43 and 0.57 but kept at 0.5 for the data
 to account for uncertainties in the Monte Carlo modelling of the efficiency 
 of the b tagging algorithm.
 This corresponds to a change in the b tagging efficiency of about
 $\pm 2\%$.
\end{description}

All uncertainties are added in quadrature to determine the total systematic
uncertainty.

\section{Conclusion}
$\dss\to\gamma\ds$ decays are selected in the invariant
mass distribution of the photon and the $\ds$ meson.
The branching ratio of $\ds$ meson decays into $\tau\nu_\tau$ has been
measured to be
$$\mbox{BR}(\dst)=\resultbr\, ,$$
in good agreement with the only other direct measurement but with a slightly
smaller uncertainty~\cite{bib-l3ds}.
From this measurement the $\ds$ decay constant can be derived
using Eq.~\ref{eq1} and the values in Table~\ref{br-table} to be
$$\fds =\resultf$$
consistent with theoretical 
predictions~[1-3] for $\fds$
and with the world average $\fds=280\pm48$~MeV~\cite{bib-pdg}.

\section*{Acknowledgements}
We particularly wish to thank the SL Division for the efficient operation
of the LEP accelerator at all energies
 and for their continuing close cooperation with
our experimental group.  We thank our colleagues from CEA, DAPNIA/SPP,
CE-Saclay for their efforts over the years on the time-of-flight and trigger
systems which we continue to use.  In addition to the support staff at our own
institutions we are pleased to acknowledge the  \\
Department of Energy, USA, \\
National Science Foundation, USA, \\
Particle Physics and Astronomy Research Council, UK, \\
Natural Sciences and Engineering Research Council, Canada, \\
Israel Science Foundation, administered by the Israel
Academy of Science and Humanities, \\
Minerva Gesellschaft, \\
Benoziyo Center for High Energy Physics,\\
Japanese Ministry of Education, Science and Culture (the
Monbusho) and a grant under the Monbusho International
Science Research Program,\\
Japanese Society for the Promotion of Science (JSPS),\\
German Israeli Bi-national Science Foundation (GIF), \\
Bundesministerium f\"ur Bildung und Forschung, Germany, \\
National Research Council of Canada, \\
Research Corporation, USA,\\
Hungarian Foundation for Scientific Research, OTKA T-029328, 
T023793 and OTKA F-023259.\\


\newpage
\begin{table}[htb]
\begin{center}
\begin{tabular}{l|l}
\hfill BR($\dst$)\hspace*{\fill}  &\hfill $\fds$ \hfill\rule[-3mm]{0mm}{8mm} \\
\hline
& $G_{\rm F}=(1.16639 \pm 0.00001)\times 10^{-5}\, {\rm GeV}^{-2}$ \\
$\mbox{BR}(\dss\to\gamma\ds)=0.942 \pm 0.025 $ &
$|V_{\rm cs}|=0.9891 \pm 0.016$ \\
$\mbox{BR}(\tau^-\to e^-\bar{\nu}_e\nu_\tau)=0.1783\pm 0.0006$ &
$m_\tau=1.77703 \pm 0.00003$ GeV \\
$\mbox{BR}(\tau^-\to \mu^-\bar{\nu}_\mu\nu_\tau)=0.1737\pm 0.0007$ &
$\mds=1.9686 \pm 0.0006$ GeV \\
& $\tds=(0.496 \pm 0.01)\times 10^{-12}\, {\rm s}$ \\
\end{tabular}
\end{center}
\caption{\label{br-table}%
External values used in the calculation of BR($\dst$) and 
$\fds$~\protect\cite{bib-pdg}.
}
\end{table}
\begin{table}[hbt]
\begin{center}
\begin{tabular}{l|c}
Source & $\Delta$ BR/BR $(\%)$\\
\hline
External Sources:     &         \\
$\pz f(c\to\ds)$      & $20.7$  \\
$\pz P_V(\dssf,\dsf)$      & $\pz7.8$  \\
$\pz R_{\mbox{c}}$    & $\pz2.9$\\
$\pz \mbox{BR}(\dss\to\gamma \ds)$ & $\pz2.7$\\
$\pz \mbox{BR}(\tmu)$ & $\pz0.2$ \\
$\pz \mbox{BR}(\tel)$ & $\pz0.1$\\
Monte Carlo: & \\
$\pz$background statistics    & $11.5$  \\
$\pz\epsilon (\dst)$ & $\pz2.3$\\
$\pz\epsilon (\dsm)$ & $\pz0.8$\\
$\pz$background rate & $\pz0.3$\\
$\pz$background composition     &  $\pz1.6$ \\
Fragmentation & $\pz1.3$\\
Lepton spectrum & $\pz0.4$\\
Detector resolution: &  \\
$\pz$ Tracking resolution & $\pz4.7$\\
$\pz$ Photon energy & $\pz6.9$\\
Lepton identification: &  \\
$\pz$ electrons & $\pz2.0$\\
$\pz$ muons     & $\pz4.6$\\
Neural network inputs & $\pz$8.3 \\
b tagging & $\pz3.0$ \\
\hline
total & $28.6$
\end{tabular}
\end{center}
\caption{\label{tab-syst}Summary of relative systematic uncertainties on 
$\mbox{BR}(\dst)$.}
\end{table}

\newpage

\end{document}